\documentclass[]{interact}
\usepackage[dvipsnames]{xcolor}
\usepackage[colorlinks=true,bookmarks=false,linkcolor=NavyBlue,urlcolor=NavyBlue,citecolor=NavyBlue,breaklinks]{hyperref}
\usepackage{amsmath}
\usepackage{amsfonts}
\usepackage{graphicx}
\usepackage[numbers,sort&compress,merge]{natbib}

\usepackage{braket}

\newcommand{\parti}[2]{\frac{\partial #1}{\partial #2}}

\newcommand{\diff}[2]{\frac{d #1}{d #2}}
\newcommand{\difft}[2]{\frac{d^2 #1}{d #2^2}}

\newcommand{\bk}[1]{\left(#1\right)}
\newcommand{\Bk}[1]{\left[#1\right]}
\newcommand{\BK}[1]{\left\{#1\right\}}

\newcommand{\expect}{\mathbb E}

\begin{document}

\title{Conservative classical and quantum resolution limits for
  incoherent imaging}

\author{ \name{Mankei Tsang\textsuperscript{a,b}}
  \affil{\textsuperscript{a}Department of Electrical and Computer
    Engineering, National University of Singapore, 4 Engineering Drive
    3, Singapore 117583; \textsuperscript{b}Department of Physics,
    National University of Singapore, 2 Science Drive 3, Singapore
    117551 } }



\maketitle

\begin{abstract}
  I propose classical and quantum limits to the statistical resolution
  of two incoherent optical point sources from the perspective of
  minimax parameter estimation. Unlike earlier results based on the
  Cram\'er-Rao bound, the limits proposed here, based on the
  worst-case error criterion and a Bayesian version of the
  Cram\'er-Rao bound, are valid for any biased or unbiased estimator
  and obey photon-number scalings that are consistent with the
  behaviors of actual estimators.  These results prove that, from the
  minimax perspective, the spatial-mode demultiplexing (SPADE)
  measurement scheme recently proposed by Tsang, Nair, and Lu
  [Phys.~Rev.~X \textbf{6}, 031033 (2016)] remains superior to direct
  imaging for sufficiently high photon numbers.
\end{abstract}

\section{Introduction}
In recent years, the Cram\'er-Rao bound (CRB) \cite{vantrees} has
become the standard measure of resolution for incoherent imaging
\cite{farrell66,helstrom70,zmuidzinas03,deschout,small14,chao16,diezmann17},
especially in fluorescence microscopy
\cite{deschout,small14,chao16,diezmann17}, where photon shot noise has
become the dominant noise source and a statistical treatment of
resolution has become essential. An often overlooked caveat of the
bound is its assumption of unbiased estimators \cite{vantrees}.
Although the fluorescence-microscopy community has embraced the
unbiased condition in principle
\cite{deschout,small14,chao16,diezmann17}, many widely used estimators
in modern statistics, including the historic maximum-likelihood (ML)
estimator \cite{vantrees,huang11} and the shrinkage estimators in
compressed sensing \cite{candes_oracle,zhu12}, can be biased and
violate the CRB, while statisticians have discovered many
counterexamples in which the unbiased condition gives rise to silly
results \cite{cox79,romano86,jaynes,maceachern93}. These issues
threaten to undermine a large body of work that proclaims the CRB as a
fundamental limit.

To remove the unbiased condition, here I propose the use of a Bayesian
CRB (BCRB) to derive new classical and quantum resolution limits to
incoherent imaging.  Following the seminal work in
Refs.~\cite{bettens,vanaert,ram}, I focus on the problem of
estimating the separation between two incoherent sources, such as
stars and fluorophores, in the presence of photon shot noise. The CRB
for this task via direct imaging has previously been proposed as a
fundamental resolution measure to supersede Rayleigh's criterion
\cite{bettens,vanaert,ram}. In a recent breakthrough, we have
discovered new measurement schemes that can significantly improve upon
direct imaging and reach the fundamental quantum limit in terms of the
Fisher information \cite{tnl,tnl2,sliver,nair_tsang16,ant};
experimental demonstrations \cite{tang16,yang16,tham16,paur16} and
further theoretical advances
\cite{tsang16c,lu16,rehacek16,lupo,krovi16,kerviche17} have since been
reported. Numerical simulations and experimental data analysis,
however, have found that sensible estimators, including the ML
estimator, are biased and violate the CRB
\cite{tnl,sliver,ant,tang16,tham16}, calling into question the
generality of the claims. To overcome this issue, here I adopt a BCRB
\cite{schutzenberger57,vantrees} that is valid for any biased or
unbiased estimator, closing the loophole of the original CRB.  By
applying the bound to a conservative worst-case error measure, I
propose new classical and quantum limits to the separation estimation
problem, and also prove that the spatial-mode demultiplexing (SPADE)
scheme proposed in Ref.~\cite{tnl} remains superior to direct
imaging for sufficiently high photon numbers and can reach the quantum
limit.

\section{\label{sec_limits}Conservative limits to two-source
  separation estimation}
The application of interest here is the estimation of the separation
between two incoherent optical point sources from far-field
measurements with Poisson noise
\cite{bettens,vanaert,ram,tnl,tnl2,sliver,nair_tsang16,ant,tang16,yang16,tham16,paur16}. Each
source produces an optical field on the image plane that is blurred by
diffraction.  Direct imaging, which measures the intensity
distribution on the image plane, performs poorly when Rayleigh's
criterion is violated \cite{bettens,vanaert,ram}.  Recently, we have
invented an alternative measurement scheme called SPADE that performs
further coherent processing on the image-plane field before photon
counting and possesses a much higher Fisher information for
sub-Rayleigh separations \cite{tnl,tnl2}.  For the uninitiated
readers, Appendix~\ref{app_crb} briefly reviews the prior results on the
use of the CRB for this separation estimation problem.

Given an observation random variable $y$, the expectation operation
$\expect_\theta$ conditioned on the unknown parameter $\theta$, and an
estimator $\check\theta(y)$, the mean-square error (MSE) is defined as
\cite{vantrees}
\begin{align}
\textrm{MSE}(\theta) &\equiv \expect_\theta\bk{\check\theta-\theta}^2.
\label{MSE}
\end{align}
If biased estimators are allowed, $\textrm{MSE}(\theta)$ can be
arbitrarily low, and no meaningful lower bound on the whole error
function can be derived. To see this point, consider a deterministic
estimator $\check\theta(y) = \theta_0$, which leads to zero error when
$\theta$ happens to be $\theta_0$. This means that
$\textrm{MSE}(\theta)$ at any specific parameter value is a poor
indicator of the overall uncertainty of the experiment if arbitrary
estimators are permitted.  While this example seems artificial, modern
statistics research has discovered many biased estimators that can
beat the CRB in useful scenarios \cite{candes_oracle,kay_eldar}.  To
derive limits that are also valid for biased estimators, here I adopt
the minimax paradigm \cite{cox79,candes_oracle,berger}, which regards
the worst-case error $\sup_\theta\textrm{MSE}(\theta)$ as the central
figure of demerit, and use the BCRB as a lower bound on the worst-case
error; see Appendix~\ref{app_bcrb} for a more detailed discussion of
the approach.

For simplicity, I assume one-dimensional imaging with $L$ detected
photons and a Gaussian point-spread function, the intensity of which
has a standard deviation $\sigma$ with respect to the object-plane
dimension; generalizations for two dimensions, uncertain photon
numbers, and other point-spread functions should give similar results
\cite{tnl,tnl2,ant,rehacek16,kerviche17}.  For the estimation of the
separation parameter $\theta$ with SPADE, Appendix~\ref{app_proof1}
shows that
\begin{align}
\sup_\theta \textrm{MSE}^{(\textrm{SPADE})}(\theta) \ge \frac{4\sigma^2}{L}.
\label{Rspade}
\end{align}
The right-hand side of Eq.~(\ref{Rspade}) also serves as a fundamental
quantum limit to $\sup_\theta \textrm{MSE}(\theta)$ for any
measurement on the image plane.  Equation~(\ref{Rspade}) has the
appearance of the CRB, but note that this is a lower bound on the
worst-case error only, not the whole function $\textrm{MSE}(\theta)$,
and it is valid for any biased or unbiased estimator. Further
theoretical and numerical analysis, to be presented in
Sec.~\ref{sec_attain}, shows that Eq.~(\ref{Rspade}) is reasonably
tight.

For direct imaging, on the other hand, Appendix~\ref{app_proof1} shows
that
\begin{align}
  \sup_\theta \textrm{MSE}^{(\textrm{direct})}(\theta) &\ge 
\frac{\sqrt{2}\sigma^2}{3\sqrt{L}}.
\label{Rdirect}
\end{align}
The $1/\sqrt{L}$ scaling implies that the direct-imaging error must
drop more slowly with increasing photon numbers than the optimal $1/L$
scaling. Given the numerous bounds involved in deriving
Eq.~(\ref{Rdirect}), there is no reason to expect it to be tight,
although the $1/\sqrt{L}$ scaling was observed in the numerical
analysis of the ML estimator in Sec.~\ref{sec_attain} and also by Tham
\textit{et al.}\ for their direct-imaging estimator \cite{tham16},
suggesting that at least the scaling is attainable. Most importantly,
Eq.~(\ref{Rdirect}) holds for any biased or unbiased estimator,
closing a crucial loophole in Refs.~\cite{bettens,vanaert,ram}.

\section{\label{sec_attain}Attaining the limits}
\begin{figure}[htbp!]
\centerline{\includegraphics[width=\textwidth]{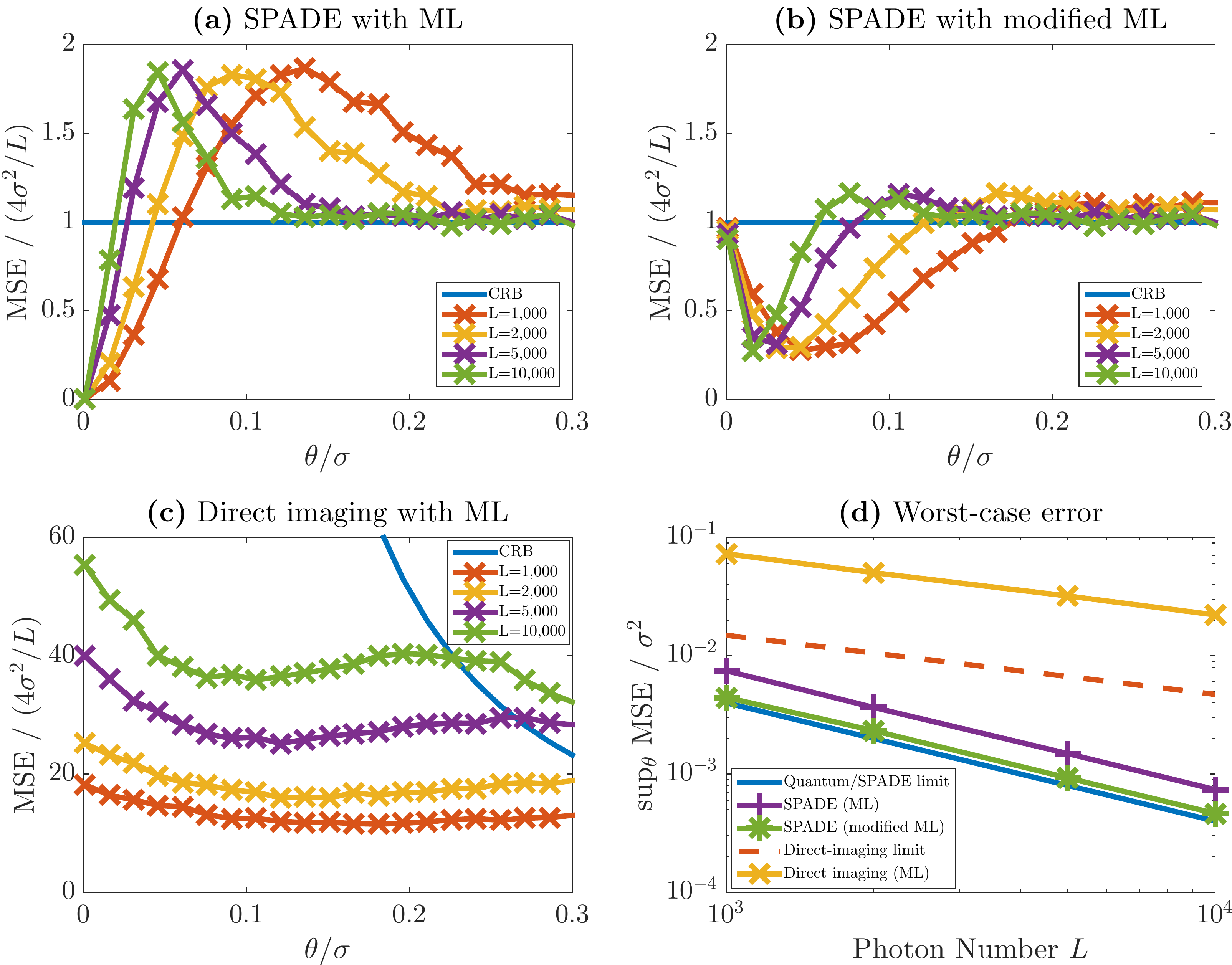}}
\caption{\label{minimax_errors}Numerically simulated MSE versus true
  separation $\theta/\sigma$ for (a) SPADE with the ML estimator, (b)
  SPADE with a modified ML estimator to reduce
  $\sup_\theta\textrm{MSE}$, and (c) direct imaging with ML.  The
  intensity of the point-spread function is assumed to be
  one-dimensional and Gaussian with standard devitation $\sigma$, and
  $L$ is the detected photon number. Note that each MSE in plots
  (a)--(c) is multiplied by $L/(4\sigma^2)$ for comparison with the
  CRBs. Each data point is an average of the errors computed from
  $5,000$ simulated measurements.  (d) Log-log plots of the
  numerically computed $\sup_\theta$ MSE for various estimators versus
  $L$, compared with the BCRBs given by Eqs.~(\ref{Rspade}) and
  (\ref{Rdirect}). The lines connecting the data points are guides for
  eyes.}
\end{figure}

For SPADE with the ML estimator, Appendix~\ref{app_proof2} proves that
\begin{align}
\textrm{MSE}^{(\textrm{SPADE, ML})}(\theta) &\le \frac{16\sigma^2}{L}.
\label{upperbound}
\end{align}
This is a guarantee of the SPADE performance for any photon number and
proves that the BCRB given by Eq.~(\ref{Rspade}) is tight up to a
prefactor of 4.  It is a more conclusive result than the
asymptotic argument or the simulations reported in Ref.~\cite{tnl}.
Figure~\ref{minimax_errors}(a) plots the simulated errors of the ML
estimator against the separation $\theta/\sigma$, showing that the
errors can actually stay less than twice the limit.  A modified ML
estimator described in Appendix~\ref{app_mml} can reduce the
worst-case error even further; Fig.~\ref{minimax_errors}(b) plots the
simulated errors of this estimator, demonstrating its near-optimality
with respect to the worst-case error criterion.

For direct imaging, Fig.~\ref{minimax_errors}(c) shows that the
simulated errors of the ML estimator violate the CRB for small
$\theta$, as also reported earlier by Ref.~\cite{huang11}, but the
important point here is that the worst-case errors must stay above the
BCRB given by Eq.~(\ref{Rdirect}). Figure~\ref{minimax_errors}(d)
compares the simulated worst-case errors of the various estimators
with the BCRBs given by Eqs.~(\ref{Rspade}) and (\ref{Rdirect}) as a
function of $L$ in log-log scale, demonstrating the different
photon-number scalings. The direct-imaging limit is less tight to the
ML errors, although the latter still appear to obey the same
$1/\sqrt{L}$ scaling.

\section{Conclusion}
I have presented new limits to the resolution of two incoherent
sources from the minimax perspective. These limits are valid for any
biased or unbiased estimator, closing a crucial loophole of previous
work based on the CRB, while the analytic and numerical results
presented here suggest that the limits can have attainable
photon-number scalings. The different scalings prove that SPADE can
significantly outperform direct imaging for sufficiently high photon
numbers. This conclusion is consistent with our earlier claims based
on the CRB \cite{tnl}, which is meaningful mainly in terms of
asymptotics \cite{gill95,cox79,hayashi05}. Beyond two sources, the
BCRB can also be applied to more general imaging scenarios
\cite{deschout,small14,chao16,diezmann17,huang11,zhu12,schiebinger,tsang16c};
it implies that the Fisher information remains a relevant precision
measure even if biased estimators are allowed. As the minimax approach
also underpins the theory of compressed sensing \cite{candes_oracle},
which has recently gained popularity not least in superresolution
microscopy \cite{zhu12}, the approach advocated here can provide a
bridge between the vast literature on the CRB
\cite{farrell66,helstrom70,zmuidzinas03,deschout,small14,chao16,diezmann17}
and the modern minimax statistics literature.

I have focused on the BCRB, but there are in fact more advanced
Bayesian bounds, such as the Ziv-Zakai family and the Weiss-Weinstein
family \cite{bell}, that can be much tighter for certain
problems. Quantum versions of such bounds have also been proposed
\cite{yuen_lax,gill_massar,gill_guta,twc,tsang_open,nair2012,hall_prx,qzzb,qbzzb,localization,qwwb,liu16}.
Given the need to deal with biased estimators and include prior
information in modern statistics, the Bayesian bounds are envisioned
to play a more prominent role in future classical and quantum imaging
applications.

\section*{Acknowledgements}
Discussions with Ranjith Nair, Xiao-Ming Lu, Shan Zheng Ang, Shilin
Ng, Edwin Tham, Hugo Ferretti, Aephraim Steinberg, Jaroslav Rehacek,
and Zdenek Hradil are gratefully acknowledged. This work is supported
by the Singapore National Research Foundation under NRF Grant
No.~NRF-NRFF2011-07 and the Singapore Ministry of Education Academic
Research Fund Tier 1 Project R-263-000-C06-112.


\bibliographystyle{tfp}

\bibliography{minimax_quantum_JMO}

\appendix
\section{\label{app_crb}Cram\'er-Rao bound (CRB)}
Let the probability distribution of the observation $y$ be
$P(y|\theta)$, which is assumed to be a function of the unknown scalar
parameter $\theta$; generalization for vectoral parameters is
straightforward \cite{vantrees}.  The expectation $\expect_\theta$ is
defined with respect to $P(y|\theta)$ as
$\expect_\theta f(y) \equiv \sum_y P(y|\theta) f(y)$. Given the
observation, an estimator $\check\theta(y)$ can be computed to
estimate the unknown parameter.  Its performance can be quantified by
the MSE defined by Eq.~(\ref{MSE}). For any unbiased estimator, which
is defined by the condition
\begin{align}
\expect_\theta\bk{\check\theta} &= \theta,
\label{unbiased}
\end{align}
the CRB is
\begin{align}
\textrm{MSE}(\theta) &\ge \frac{1}{J(\theta)},
\end{align}
where 
\begin{align}
  J(\theta) &\equiv 
\expect_\theta \Bk{\parti{\ln P(y|\theta)}{\theta}}^2
\end{align}
is the Fisher information. In a quantum problem, the Fisher
information is in turn upper-bounded by quantum versions of the Fisher
information \cite{hayashi}, which depend on the density operator of
the quantum object being measured \cite{hayashi,helstrom,holevo11}.

\begin{figure}[htbp!]
\centerline{\includegraphics[width=0.6\textwidth]{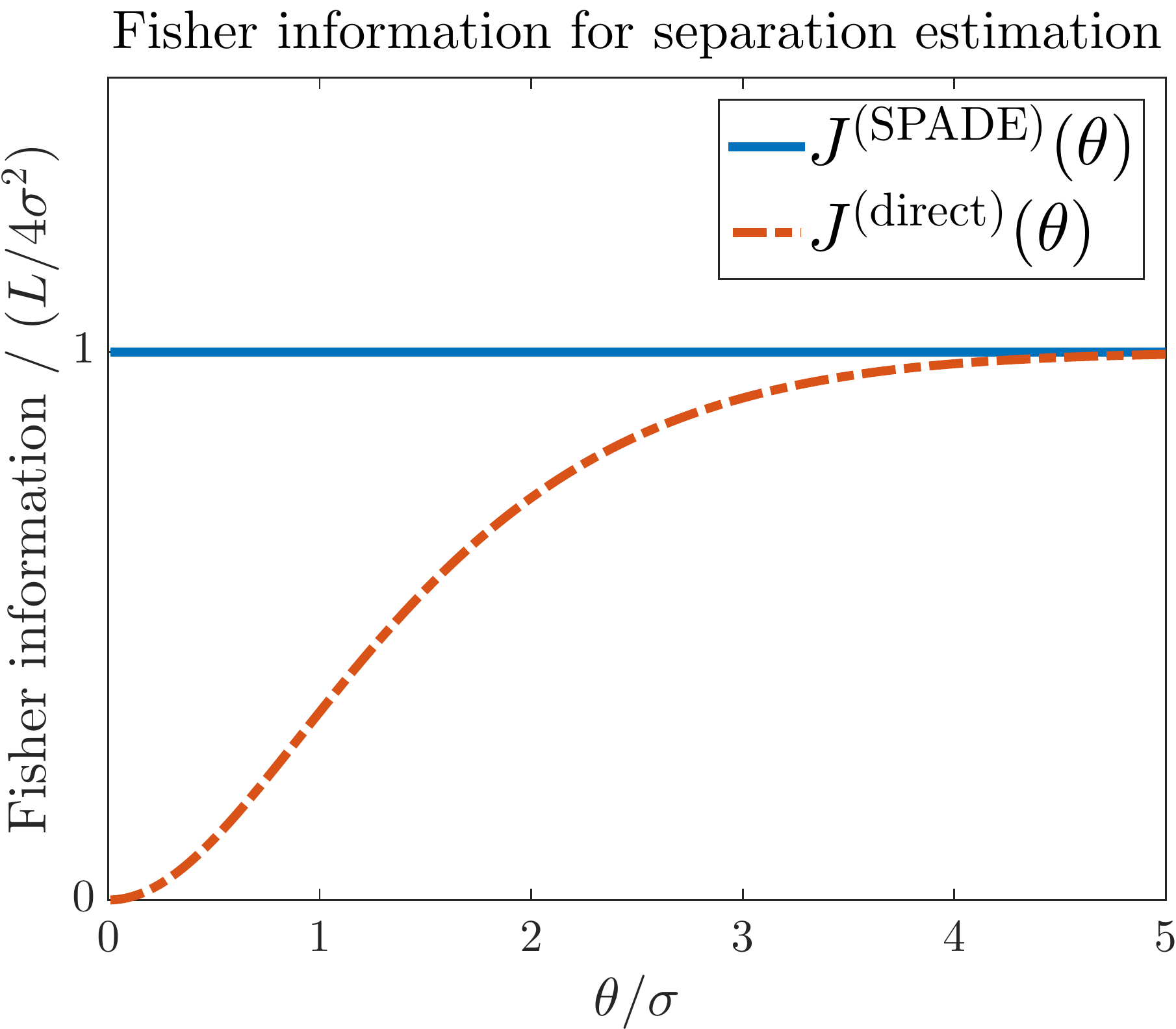}}
\caption{Fisher information for the estimation of the separation
  between two incoherent point sources with the SPADE measurement
  scheme ($J^{(\textrm{SPADE})}(\theta)$) and direct imaging
  ($J^{(\textrm{direct})}(\theta)$) versus normalized separation
  $\theta/\sigma$, where $\sigma$ is the width of the point-spread
  function and $L$ is the detected photon number. The information is
  normalized with respect to the shot-noise limit $L/(4\sigma^2)$.}
\label{fisher}
\end{figure}

Figure~\ref{fisher} plots the Fisher information for the separation
estimation problem with SPADE and direct imaging for a given photon
number $L$ \cite{tnl}. The vertical axis is normalized with respect to
the shot-noise limit $L/(4\sigma^2)$ and the horizontal axis is the
true separation $\theta$ normalized with respect to the
point-spread-function width $\sigma$.  For large separations, both
quantities approach the shot-noise limit, but the information
$J^{(\textrm{direct})}(\theta)$ for direct imaging drops to zero for
$\theta \lesssim \sigma$, a phenomenon discovered by
Refs.~\cite{bettens,vanaert,ram} and called Rayleigh's curse in our
previous work \cite{tnl}. The CRB
$\textrm{MSE}^{(\textrm{direct})}(\theta) \ge
1/J^{(\textrm{direct})}(\theta)$ suggests that the error of any
unbiased estimator must blow up for $\theta \to 0$, but it does not
rule out the possibility that biased estimators can do better.
Indeed, studies have found that the CRB for direct imaging can be
violated for small $\theta$ \cite{huang11,tham16}. The Fisher
information $J^{(\textrm{SPADE})}(\theta)$ for SPADE, meanwhile,
remains constant and is given by
\begin{align}
J^{(\textrm{SPADE})}(\theta) &= \frac{L}{4\sigma^2},
\label{Jspade}
\end{align}
which also coincides with the quantum limit to the Fisher information
for any measurement \cite{tnl}.

\section{\label{app_bcrb}Bayesian Cram\'er-Rao bound (BCRB)}
To investigate fundamental limits to the worst-case error of any
estimator, one can take advantage of the fact that the error is always
higher than the error averaged over a prior probability density
$p(\theta)$, viz.,
\begin{align}
\sup_\theta\textrm{MSE}(\theta) \ge \int d\theta p(\theta)\textrm{MSE}(\theta)
\equiv B[p].
\end{align}
Lower bounds on $B[p]$ can then be used on the worst-case error.  In
particular, for any $p(\theta)$ that converges to zero at the
endpoints of the interval of $\theta$, the BCRB reads
\cite{schutzenberger57,vantrees,gill95}
\begin{align}
\sup_\theta\textrm{MSE}(\theta) &\ge B[p] \ge \frac{1}{K[p]},
\label{BCRB}
\\
K[p] &\equiv \int d\theta p(\theta) J(\theta) + j[p],
\label{vtinfo}
\\
j[p] &\equiv 
\int d\theta p(\theta)
\Bk{\parti{\ln p(\theta)}{\theta}}^2.
\end{align}
The BCRB is sometimes called the Van Trees inequality \cite{gill95}
but in fact first reported by Sch\"utzenberger in 1957
\cite{schutzenberger57,bell}.  The bound is
appealing for two reasons: it is valid for any estimator, not just
unbiased ones, and it depends on the Fisher information, a quantity
that has been studied extensively for many applications.  Quantum
versions of the BCRB
\cite{yuen_lax,gill_massar,gill_guta,twc,tsang_open} follow naturally
from quantum upper bounds on $J(\theta)$.

The BCRB on the worst-case error is valid for any $p(\theta)$ that
satisfies the zero boundary conditions, meaning that one can choose a
$p(\theta)$ that tightens the bound. A trick is to assume
$p(\theta) = q^2(\theta)$, such that
\begin{align}
K &= \int d\theta \BK{q^2(\theta) J(\theta) + 4\Bk{\diff{q(\theta)}{\theta}}^2},
\label{energy}
\end{align}
and the minimizing solution, subject to the normalization condition
$\int d\theta q^2(\theta) = 1$, obeys the Euler-Lagrange equation
\begin{align}
\lambda q(\theta) &=
-4\difft{q(\theta)}{\theta} + J(\theta)q(\theta),
\label{q}
\end{align}
where $\lambda$ is a Lagrange multiplier. As Eq.~(\ref{q}) has the
same form as the time-independent Schr\"odinger equation and
Eq.~(\ref{energy}) has the form of the average energy of a
wavefunction, computing the minimum $K[p]$ is equivalent to finding
the ground-state energy in the wave problem, with $J(\theta)$ playing
the role of the potential and the prior information $j[p]$ playing the
role of the average kinetic energy.

\section{\label{app_proof1}Proof of Eqs.~(\ref{Rspade}) and (\ref{Rdirect})}
For SPADE, the constant Fisher information given by
Eq.~(\ref{Jspade}) means that
$\int d\theta p(\theta) J^{(\textrm{SPADE})}(\theta) = L/(4\sigma^2)$
for any prior, and I can choose an uninformative prior with
$j[p] \to 0$ to obtain
$\inf_p K^{(\textrm{SPADE})}[p] = L/(4\sigma^2)$, which gives
Eq.~(\ref{Rspade}) via the BCRB in Eq.~(\ref{BCRB}).  A
quantum Fisher information coincides with
$J^{(\textrm{SPADE})}(\theta)$ \cite{tnl} and mandates that
$J(\theta) \le J^{(\textrm{SPADE})}(\theta)$ for any quantum
measurement \cite{hayashi}, so the Bayesian information for any
measurement obeys $K[p] \le K^{(\textrm{SPADE})}[p]$, and the
right-hand side of Eq.~(\ref{Rspade}) also serves as a
fundamental quantum limit.

Deriving a tight bound on
$\sup_\theta\textrm{MSE}^{(\textrm{direct})}(\theta)$ for direct
imaging is more nontrivial. To simplify, note that, since
$J^{(\textrm{direct})}(\theta)\propto L$, the optimal $p(\theta)$
should be highly concentrated near $\theta = 0$ for large $L$, in
which case I can use a quadratic upper bound on
$J^{(\textrm{direct})}(\theta)$ that is tight near $\theta = 0$ to
approximate it \cite{bettens}, viz.,
\begin{align}
J^{(\textrm{direct})}(\theta) &\le \frac{L\theta^2}{8\sigma^4}.
\label{Jbound}
\end{align}
With this upper bound in place of $J^{(\textrm{direct})}(\theta)$ in
Eq.~(\ref{q}) and the boundary conditions $q(0) = q(\infty) = 0$, it
is well known that the solutions of Eq.~(\ref{q}) are odd-order
Hermite-Gaussian functions. Taking the lowest order, the result for
$p(\theta) = q^2(\theta)$ is
\begin{align}
p(\theta) &= 
\sqrt{\frac{2}{\pi}}\frac{\theta^2}{w^3}
\exp\bk{-\frac{\theta^2}{2w^2}},
\quad
w^2 = \sigma^2\sqrt{\frac{8}{L}}.
\label{prior}
\end{align}
Substituting Eqs.~(\ref{Jbound}) and (\ref{prior}) into
Eq.~(\ref{vtinfo}) yields
\begin{align}
K^{(\textrm{direct})}[p] \le 
\frac{3L w^2}{8\sigma^4} + \frac{3}{w^2} = \frac{3\sqrt{L}}{\sqrt{2}\sigma^2},
\end{align}
resulting in Eq.~(\ref{Rdirect}) by virtue of the BCRB in
Eq.~(\ref{BCRB}). Alternatively, the same result can be obtained by
assuming the prior to be Eq.~(\ref{prior}) with a free hyperparameter
$w$ and then choosing the $w$ that gives the tightest bound.

\section{\label{app_proof2}Proof of Eq.~(\ref{upperbound})}
Ref.~\cite{tnl} shows that the ML estimator for SPADE can be
expressed as
\begin{align}
\check\theta^{(\textrm{SPADE, ML})} &= 4\sigma \sqrt{\frac{Y}{L}},
\end{align}
where $Y$ is a Poisson statistic summarizing the measurement outcomes,
with a mean given by
\begin{align}
\expect_\theta(Y) &=  \frac{L\theta^2}{16\sigma^2}.
\label{expectY}
\end{align}
The error becomes
\begin{align}
\textrm{MSE}^{(\textrm{SPADE, ML})}(\theta) 
&= \expect_\theta\bk{\check\theta^{(\textrm{SPADE, ML})}-\theta}^2
\\
&= 2\theta\Bk{\theta - \frac{4\sigma}{\sqrt{L}}
\expect_\theta\bk{\sqrt{Y}}}.
\label{mse_expect}
\end{align}
Using the inequality
\begin{align}
\sqrt{X} &\ge 1 + \frac{1}{2}
\bk{X-1} - \frac{1}{2}\bk{X-1}^2
\textrm{ for }
X \ge 0,
\label{ineq}
\end{align}
which follows from $(\sqrt{X}-1)^2(\sqrt{X}+2)\sqrt{X}/2 \ge 0$, setting
$X = Y/\expect_\theta(Y)$ in Eq.~(\ref{ineq}), and taking the
expectation on both sides, I obtain
\begin{align}
\expect_\theta\bk{\sqrt{Y}} &\ge \sqrt{\expect_\theta(Y)}-
\frac{1}{2\sqrt{\expect_\theta(Y)}}.
\end{align}
Combining this inequality with Eqs.~(\ref{expectY}) and
(\ref{mse_expect}) leads to Eq.~(\ref{upperbound}).

\section{\label{app_mml}Modified maximum-likelihood (ML) estimator}
The vanishing error at $\theta = 0$ in Fig.~\ref{minimax_errors}(a)
implies that the ML estimator is heavily biased towards that value.
This suggests that the worst-case error, which happens at larger
$\theta$, can be improved by introducing a more positive bias.  As
discovered empirically, an improvement can be obtained simply by
setting the ML estimator for $Y = 0$ to $2\sigma/\sqrt{L}$ rather than
$0$. In other words,
\begin{align}
\check\theta &= 
\left\{\begin{array}{ll}
2\sigma/\sqrt{L}, & Y = 0,\\
4\sigma \sqrt{Y/L}, & Y \neq 0.
\end{array}
\right.
\end{align}
Figures~\ref{minimax_errors}(b) and (d) plot the simulated errors of
this estimator, demonstrating a noticeable improvement over the ML
estimator.


\end{document}